\documentclass[pra,twocolumn,showpacs,superscriptaddress,floatfix]{revtex4}
\usepackage[dvips]{graphicx}
\usepackage{txfonts}
\begin{document}
\title{%
    Experimental demonstration of entanglement assisted coding\\
    using a two-mode squeezed vacuum state}
\author{Jun Mizuno}
\affiliation{%
    National Institute of Information and Communications Technology (NICT)
    4--2--1 Nukuikita, Koganei 184--8795, Japan}
\affiliation{%
    CREST, Japan Science and Technology Corporation,
    3--13-11 Shibuya, Shibuya-ku, Tokyo 150--0002, Japan}
\author{Kentaro Wakui}
\affiliation{%
    Department of Applied Physics,
    The University of Tokyo,
    7--3--1 Hongo, Bunkyo-ku, Tokyo 113--8656, Japan}
\author{Akira Furusawa}
\affiliation{%
    Department of Applied Physics,
    The University of Tokyo,
    7--3--1 Hongo, Bunkyo-ku, Tokyo 113--8656, Japan}
\author{Masahide Sasaki}
\affiliation{%
    National Institute of Information and Communications Technology (NICT)
    4--2--1 Nukuikita, Koganei 184--8795, Japan}
\affiliation{%
    CREST, Japan Science and Technology Corporation,
    3--13-11 Shibuya, Shibuya-ku, Tokyo 150--0002, Japan}
\email{e-mail:psasaki@nict.go.jp}
\begin{abstract}
We have experimentally realized the scheme initially proposed
as quantum dense coding with continuous variables
[Ban, J.~Opt.~B \textbf{1}, L9 (1999), and
 Braunstein and Kimble, \pra\textbf{61}, 042302 (2000)].
In our experiment,
a pair of EPR (Einstein-Podolski-Rosen) beams is generated
from two independent squeezed vacua.
After adding two-quadrature signal to one of the EPR beams,
two squeezed beams that contain the signal were recovered.
Although our squeezing level is not sufficient to demonstrate
the channel capacity gain over the Holevo limit of
a single-mode channel without entanglement,
our channel is superior to conventional channels such as
coherent and squeezing channels.
% our channel capacity exceeds those of conventional channels
% such as the coherent channel or squeezing channel.
% the channel capacity gain over the classical limit,
In addition,
optical addition and subtraction processes demonstrated
are elementary operations of universal quantum information processing
on continuous variables.
\end{abstract}
%% 03.67.-a Quantum information
% 03.67.Hk Quantum communication
%% 03.65.Ta Foundations of quantum mechanics; measurement theory  
%% 42.50.-p Quantum optics
% 42.50.Dv Nonclassical states of the electromagnetic field, including
%          entangled photon states; quantum state engineering and measurements
%
\date{13.\ May,\ 2004}
%\date{\today}
%
\maketitle
%
%%%%%%%%%%%%%%%%%%%%%%%%%%%%%%%%%%%%%%%%%%%%%%%%%%%%%%%%%%%%
%
\section{Introduction}
The entanglement of the EPR beams generated by superimposing two
independent squeezed beams\,%
\cite{FK98}
is utilized in continuous-variable (CV) quantum information
experiments, such as quantum teleportation\,%
\cite{FK98,ZGCLK03,BTBSRBSL03}
and quantum dense coding\,%
\cite{Ban99,Ban00,BK00,LJZXP02,RH02}.
The latter scheme utilizes this entanglement to enhance the
channel capacity of a communications channel by reducing
the vacuum noises in both of two quadratures of a signal.
This can also be applied to
sub-shot-noise sensing and cryptography.

The principle of CV dense coding was first demonstrated
by Li et~al.\,%
\cite{LJZXP02}
in a form somewhat simplified from the original proposal\,%
\cite{BK00}.
% ,
% that is, using bright EPR beams
% and direct detection of photocurrents
% without any additional local oscillators (LOs),
% called self homodyning.
They used bright EPR beams and detected the photocurrents directly
without any additional local oscillators (LOs)
in the scheme they called self homodyning.
The simplicity of their scheme can be
an advantage for implementation of, e.g.,
CV quantum cryptography.
In terms of channel capacity in the power constrained scenario,
however, their scheme cannot be efficient
% since most of the light power in the channel is squandered
% for self-homodyning (playing the role of LO)
since self homodyning requires the bright carrier component
(that plays the role of LO) sent through the channel.
This carrier component occupies the major fraction of
the limited power in the channel
without contributing to the signal.
We will return to this point later in the discussion
(Section~\ref{sc:rd}).

In this paper,
we describe our experiment that is performed
in the way initially proposed by Braunstein and Kimble\,%
\cite{BK00}.
Two independent squeezed vacua are used as the EPR source and
separate LOs are used in homodyne detection.
The necessity of independent LOs
makes our setup elaborate, but
this allows us broader and more flexible operations 
on CV\,%
\cite{SB98a,SB98b,LloydBraunstein99,GKP01,BS02}.
In addition, the use of squeezed vacua is 
valuable in sensing of a system
whose state may be disturbed by a bright light probe.

%%%%%%%%%%%%%%%%%%%%%%%%%%%%%%%%%%%%%%%%%%%%%%%%%%%%%%%%%%%%
%
\section{Principle}
\begin{figure}[!htb]
\includegraphics[width=\hsize]{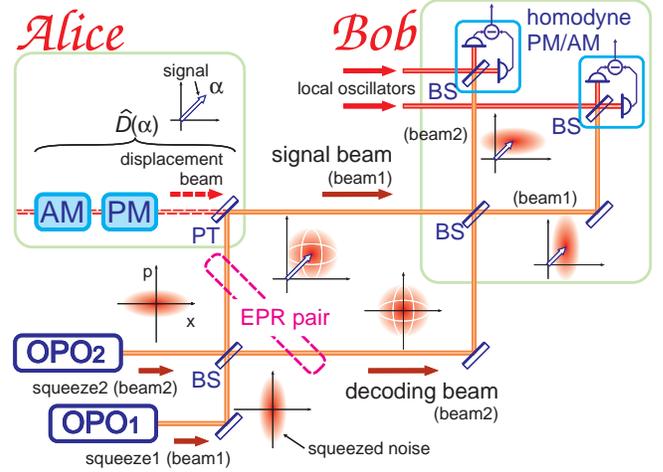}%
\caption{\label{fg:prin}
Schematic diagram of dense coding experiment using
squeezed vacuum states
(see text for details).
BS: 50\,:\,50 beamsplitter and
PT: partially transmitting mirror.
}
\end{figure}
The actual experiment shown in Fig.~\ref{fg:prin}
is carried out using optical sideband modes,
although the principle below is described in the carrier mode.
The squeezed states show nonclassical correlations
between upper and lower sidebands,
and the two-quadrature displacement corresponds to the amplitude
and phase modulations (AM, PM) of the displacement beam.
Homodyne detection converts the amplitudes of two optical sidebands
to the photocurrents of the frequency near DC,
which is then measured electrically.

The squeezed vacuum is generated by
an optical parametric oscillator (OPO)
operated below threshold.
In our setup, 
operations of OPO1 and OPO2 are described by
\begin{eqnarray}
\label{eq:sqza}
\hat{x}_1=\exp(-r)\,\hat{x}^{(0)}_1
\,, &&
\hat{p}_1=\exp(r)\,\hat{p}^{(0)}_1
\quad\mbox{and}\\
\label{eq:sqzb}
\hat{x}_2=\exp(r)\,\hat{x}^{(0)}_2
\,, &&
\hat{p}_2=\exp(-r)\,\hat{p}^{(0)}_2
\,,
\end{eqnarray}
respectively,
where $r$ (${>}0$) is the squeezing parameter,
$\hat{x}_k$ and $\hat{p}_k$ $(k=1,2)$ are
the canonically conjugate operators
($[\hat{x}_k,\hat{p}_k]=i$)
for the two orthogonal quadratures in phase space,
which we refer to as $x$- and $p$-quadratures.
The superscript~${(0)}$ is for the input field
which is assumed to be in the vacuum state,
i.e.\
$\big\langle\hat x^{(0)}_k\big\rangle
= \big\langle\hat p^{(0)}_k\big\rangle = 0$
and
$\big\langle[\Delta\hat x^{(0)}_k]^2\big\rangle
= \big\langle[\Delta\hat p^{(0)}_k]^2\big\rangle = 1/2$\,.

The output fields from the OPOs are combined by
a 50\,:\,50 beamsplitter to compose a pair of EPR beams.
They are described by the operators
\begin{eqnarray}
\label{eq:epra}
\hat{X}_1=\frac{\hat{x}_1+\hat{x}_2}{\sqrt2}
\,, &&
\hat{P}_1=\frac{\hat{p}_1+\hat{p}_2}{\sqrt2}
\quad\mbox{and}\\
\label{eq:eprb}
\hat{X}_2=\frac{\hat{x}_2-\hat{x}_1}{\sqrt2}
\,, &&
\hat{P}_2=\frac{\hat{p}_2-\hat{p}_1}{\sqrt2}
\,.
\end{eqnarray}
When each EPR beam is detected separately,
we observe the so-called \emph{EPR noise}
which is larger than the shot noise level.

At Alice's site,
the beam~1 $(\hat{X}_1,\hat{P}_1)$
experiences a displacement described by the operator $\hat D(\alpha)$
where the complex number $\alpha$ represents a two-quadrature signal.
This displacement is achieved by using
a partially transmitting mirror
% ($T_\mathrm{PT}\approx 1\,\%$ in our experiment)
which reflects the EPR beam with a small loss
and adds a small fraction of the displacement beam to realize
the displacement by~$\alpha$.
This beam~1 after displacement is the signal beam
that is sent through the channel.

At Bob's site, the signal beam
is combined with the other beam of the EPR pair
(the beam 2, the \emph{decoding beam})
via a 50\,:\,50 beamsplitter.
The resulting output state is a separable product of
two displaced squeezed states.
They are then detected by two balanced homodyne detectors
to extract the signal in $x$- and $p$-quadratures
of the beams 1 and~2, respectively.
The measurement results exhibit the signal in two quadratures
\begin{eqnarray}
\label{eq:dca}
\left\langle \frac{\hat{X}_1-\hat{X}_2}{\sqrt{2}} \right\rangle
= \frac{\Re\{ \alpha \}}{\sqrt2}
\,,\qquad
\left\langle \frac{\hat{P}_1+\hat{P}_2}{\sqrt{2}} \right\rangle
= \frac{\Im\{ \alpha \}}{\sqrt2}
\end{eqnarray}
% The measurement results exhibit 
with variances smaller than
the shot noise limit ($r=0$) simultaneously,
\begin{eqnarray}
&&
\left\langle\,
\left[\Delta\!\left(\frac{\hat{X}_1-\hat{X}_2}{\sqrt{2}}\right)\right]^{2}\,
\right\rangle
=
\left\langle\,
\left[\Delta\!\left(\frac{\hat{P}_1+\hat{P}_2}{\sqrt{2}}\right)\right]^{2}\,
\right\rangle
\nonumber\\
\label{eq:dcb}
&=& \frac{1}{2}\exp(-2r)
\rightarrow 0
\qquad\mbox{as $r\to\infty$\,,}
\end{eqnarray}
by the virtue of the entanglement of EPR beams
($[\,\hat{X}_1-\hat{X}_2,\hat{P}_1+\hat{P}_2\,]=0$).

\begin{figure}
\includegraphics[width=\hsize]{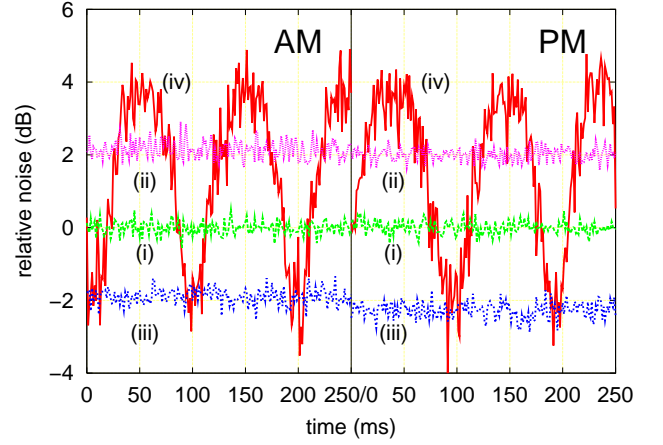}
\caption{\label{fg:tm}
Outputs from the two balanced homodyne detectors in the time domain;
(i)\,shot noise, (ii)\,EPR noise, (iii)\,squeezed beam with LO locked,
and (iv)\,squeezed beam.
All measurements are carried out without any signal ($\alpha=0$)
and with the LO scanning except for the plot~(iii).
The vertical scale is normalized by the shot noise level.
Measurement parameters are\,:
center frequency 1.1\,MHz, span 0\,Hz, resolution bandwidth 30\,kHz,
video bandwidth 0.3\,kHz,
and plots (i)---(iii) are averaged for 10 times.
}
\end{figure}
%
%%%%%%%%%%%%%%%%%%%%%%%%%%%%%%%%%%%%%%%%%%%%%%%%%%%%%%%%%%%%
%
\section{Experiment}
In our experiment,
a Ti:sapphire laser (Coherent Scotland, MBR--110)
excited by a 10\,W, 532\,nm DPSS laser (Coherent, Verdi--10)
is used as the source of $\approx$\,1.3\,W
fundamental wavelength (860\,nm).
Major fraction (90\,\%) of the beam is used to generate
second harmonic wavelength (430\,nm) of $\approx$\,120\,mW
by a frequency doubler (Coherent Scotland, MBD--200).
The second harmonic
beam is used to generate two squeezed beams
at fundamental wavelength
through parametric down-conversion processes (see below).
The rest of the fundamental beam is used as
a signal (i.e.\ displacement beam),
as LOs for balanced homodyne detection,
and also as probe beams for various control purposes.

For optical parametric down-conversion,
a 10\,mm long potassium niobate (KNbO$_3$) crystal with
temperature control for noncritical phase matching
is employed as nonlinear optical material\,%
\cite{PCK92}.
It is placed around the beam waist (with a beam radius 20\,$\mu$m)
in the cavity composed of four mirrors,
two flat and two concave (curvature radius 50\,mm) ones,
which form a bow-tie configuration
with a round-trip length of $\approx$\,500\,mm.
At the fundamental wavelength the mirrors are highly
reflective ($R_{860}>99.95\,\%$) excepting the flat output coupler
that has a transmittance of $T_{860}^{\mathrm{oc}}=10\,\%$.
All mirrors are transmissive ($T_{430}>85\,\%$)
at the second harmonic wavelength, and thus the pumping beam is
essentially single pass.

The two squeezed beams are superimposed at a 50\,:\,50
beamsplitter to compose a pair of EPR beams.
At Alice, one of them (signal beam) is superimposed at
a partially transmitting mirror 
($T_\mathrm{PT}\approx 1\,\%$) % in our experiment
with a displacement beam which contains possible AM and PM signals.
The signal beam is then decoded at Bob by being superimposed
with the decoding beam at another 50\,:\,50 beamsplitter.
Each of the two resultant beams is detected by a balanced
homodyne detector %. measured by 
using photodiodes with quantum efficiencies of 99.9\,\%
(Hamamatsu, S3590 AR coated at 860\,nm).

Each squeezed beam is marked by a probe beam
($\approx$\,7\,$\mu$W after the flat output coupler)
that is taken from the fundamental light source  %%1019
and is injected in parallel to the squeezed beam through
one of the high reflectors of the OPO cavity.
Each probe beam is
phase-modulated at a frequency different from each other
(around 55\,kHz and 64\,kHz).
The relative phase to compose the EPR beams is controlled
referring this probe beam.
Similarly, other interference phases
(to superimpose the displacement beam, to decode at Bob,
and to detect by balanced homodyne detectors)
are all controlled,
either through DC interference or via demodulating
the modulation applied to the probe beam.
These control loops keep the whole optical system
to proper operating points during measurements.
(Some of them may be scanned to investigate
the phase dependences.)

\begin{figure}
\includegraphics[width=\hsize]{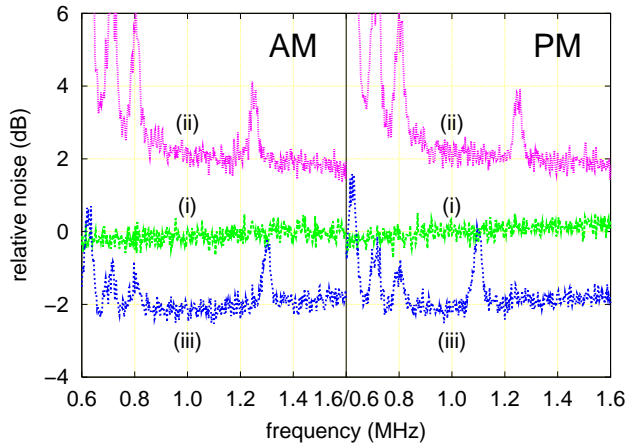}
\caption{\label{fg:fq}
Outputs from the two balanced homodyne detectors in the frequency domain;
(i)\,shot noise, (ii)\,EPR noise, and (iii)\,squeezed beam with LO locked.
All measurements are carried out with the LO scanning
except for the plot~(iii).
The vertical scale is normalized by the shot noise level
around 1.1\,MHz.
Measurement parameters are the same as those of Fig.~2,
except for the span 1\,MHz.
The peaks at 1.3\,MHz and 1.1\,MHz in plot (iii)
are simulated signals in amplitude and phase quadratures, respectively.
The peaks at 1.25\,MHz and in low frequency region
(${<}\,0.8\,\mathrm{MHz}$) are due to
the technical noises of the laser rather than the quantum noise.
Apparent frequency dependences of the shot noises are due to
the detector circuitry.
}
\end{figure}
%
%%%%%%%%%%%%%%%%%%%%%%%%%%%%%%%%%%%%%%%%%%%%%%%%%%%%%%%%%%%%
%
\section{Results and discussion}
\label{sc:rd}
The experimental results %,
% measured by photodiodes with quantum efficiencies of 99.9\,\%
% (Hamamatsu, S3590 AR coated at 860\,nm),
are shown in Figs.\,2 and~3.
The former shows the noise power around 1.1\,MHz in the time domain
and the latter shows the noise power spectra in the frequency domain,
both in two quadratures.
(See their captions for the measurement parameters.)
As can be seen in the figures,
the level of squeezing is about
2.0\,dB ($r\approx0.23$) in this experiment.

Fig.~\ref{fg:tm} shows various noise levels in each quadrature;
those are
the shot noise, the EPR noise
[the variances of the operators given in
Eqs.\,(\ref{eq:epra}) and (\ref{eq:eprb})],
and the recovered squeezed noises with LO scanning and with LO locked
[Eq.\,(\ref{eq:dcb})].
It can be seen that the EPR noise is larger than
the shot noise and also is phase independent.
This EPR noise can be canceled by combining with
the other EPR beam (decoding beam), resulting in
two separable squeezed states.
It is clearly confirmed
by the asymmetrically oscillating curves with steep dips
when the LO phases are scanned.
When the LO phases are locked to the minima,
the noise levels of both quadratures simultaneously
stay lower than that of the shot noise (the vacuum state).

Fig.~\ref{fg:fq}
shows essentially the same information as Fig.~2
in the frequency domain, with simulated AM and PM signals
at 1.3\,MHz and 1.1\,MHz, respectively.
The two simultaneously added signals are
clearly separated in the outputs of two homodyne detectors.
Obviously the signals that were buried in the vacuum noise
become apparent after the optical subtraction process.

The amount of information that can be transmitted by
a dense coding channel with the given level of squeezing~$r$
and the average signal photon number~$|\alpha|^2$ is given by
\cite{BK00,RH02}
\begin{equation}
\label{eq:Idc}
I_{\mathrm{DC}} = \ln \left[ 1+|\alpha|^2 \exp(2r) \right]
\,.
\end{equation}
Since squeezing requires a finite photon number of $\sinh^2 r$,
the average photon number in the signal beam is given by 
\begin{equation}
\label{eq:nav}
\bar n = \vert\alpha\vert^2 + \sinh^2 r
\,.
\end{equation}
% When comparing the capacities of various channels, 
In evaluating the capacities of various CV channels,
this $\bar{n}$ must be fixed, as was assumed in
Refs.\,[\onlinecite{BK00,RH02}].
We refer to this condition as the power constrained scenario.

\begin{figure}
\includegraphics[width=\hsize]{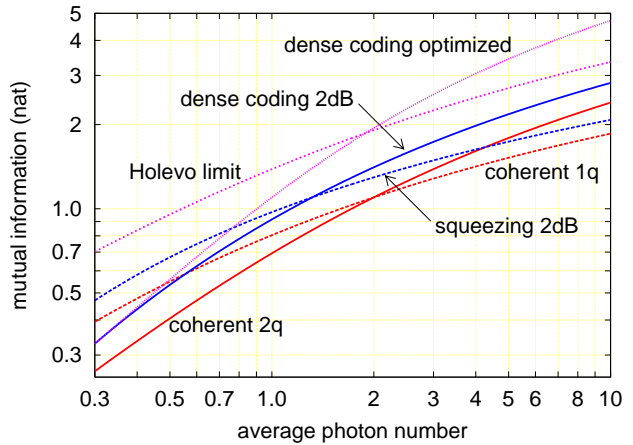}
\caption{\label{fg:ixy}
The amounts of information transmissible by various channels,
measured by mutual information,
as functions of the average photon number in the channel
[$\bar{n}$ in Eq.\,(\ref{eq:nav})].
Shown are
coherent single quadrature (coherent 1q),
coherent double quadrature (coherent 2q), 
dense coding optimized (assuming arbitrary level of squeezing available),
dense coding with 2\,dB squeezing,
homodyne with 2\,dB squeezing, and
the Holevo limit of a single-mode bosonic channel without entanglement.
The amount of information is measured in
${\mathrm{nat}} = (1/\ln 2)\,{\mathrm{bit}}
\approx 1.443\,{\mathrm{bit}}$.
}
\end{figure}
%
% A comparison of channel capacity with other channels 
% % (see Refs.\,[\onlinecite{BK00,RH02}] and references therein)
% are done in the power constrained scenario, i.e.\
% the average photon number in the signal beam
% \begin{equation}
% \label{eq:nav}
% \bar n = \vert\alpha\vert^2 + \sinh^2 r
% \end{equation}
% is limited.
% Here the first and the second terms represent the signal photons
% and squeezing photons, respectively.
Fig.~\ref{fg:ixy} depicts the amounts of information transmitted
by various channels (see the caption for details)
in the power constrained scenario, derived in literature such as
Refs.\,[\onlinecite{BK00,RH02}]
and references therein.
The dense coding channel is always superior to
a conventional coherent state channel using two quadratures.
% even with 2\,dB (or less) squeezing.
% (assuming that the squeezing level is finite and can be reduced
% arbitrarily).
With the present level of squeezing, about 2.0\,dB,
our dense coding channel has the capability of exceeding
% the conventional coherent state channel with heterodyne detection and 
the single quadrature channel using the same level of squeezing
under the condition of $\bar{n}\gtrsim 1.316$.
This condition on~$\bar{n}$
can be further reduced by the use of improved squeezing.
% under the conditions of
% $\bar{n}\gtrsim 0.146$ and $1.316$, respectively.
% These conditions will be relaxed by the use of improved squeezing.

%%1019
The ultimate performance of dense coding, optimized assuming
that an arbitrary level of squeezing is available, 
is also shown in Fig.\,\ref{fg:ixy}.
As can be seen in the figure, it exceeds the (yet-to-be-achieved)
classical limit of a bosonic channel capacity without entanglement
(the Holevo limit of a single-mode bosonic channel)
for $\bar{n}\gtrsim 1.883$ %.
with squeezing level of 6.78\,dB\,%
\cite{BK00}.
The minimum squeezing level required to beat the classical limit
for a larger $\bar{n}$ is 4.34\,dB, i.e.\ $r\ge0.5$ (see
Ref.\,[\onlinecite{RH02}]).
When such a level of squeezing becomes available,
our scheme has the potential of achieving this,
i.e., of realizing `true' quantum dense coding
as is discussed in the original proposals\,%
\cite{Ban99,BK00}.
% Once sufficiently high level of squeezing
% (4.34\,dB, i.e.\ $r\ge0.5$, see
% Ref.\,[\onlinecite{RH02}])
% is reached, our scheme has the potential of beating
% the (yet-to-be-achieved) classical limit of a bosonic channel capacity
% without entanglement
% (the Holevo limit of a single-mode bosonic channel),
% realizing `true' quantum dense coding
% as is discussed in the original proposals\,%
% \cite{Ban99,BK00}.

%%1019
This should be contrasted with the experiment in
Ref.\,[\onlinecite{LJZXP02}], 
which inevitably requires a bright component in the channel
for self homodyning.
% where a bright beam for self-homodyning is required
% in the channel.
This bright component must be taken into account as an additional term
in Eq.\,(\ref{eq:nav}) for a comparison in the power constrained scenario.
Then the information gain attained by dense coding will be diminished.
To maximize the channel capacity within a fixed power in the channel,
it is necessary to prepare separate local oscillators at the receiver
and to preserve the photons in the channel for signals.
%
% When this bright beam is taken into account as an additional term
% in Eq.(6), it will diminish the capacity gain attained by dense coding.
% For the purpose of channel capacity enhancement,
% it is much more efficient to spend more photons in the channel for signals
% and to prepare separate local oscillators at the receiver.

%%1019
Our experiment also requires further improvements, in addition
to the squeezing level, to realize true quantum dense coding.
The excess noise at antisqueezing [peaks of plot (iv) in
Fig.\,\ref{fg:tm}] contains extra photons
(assuming Gaussian noise, see
Ref.\,[\onlinecite{RH02}]) of
$ %\begin{equation}
\bar{n}_\mathrm{ex} = [\exp(r_+)-\exp(r)]/4
$ %\end{equation}
where $r$ and $r_+\,({>}0)$ are the squeezing and antisqueezing
levels, respectively.
From the uncertainty principle $r_+ \ge r$ is required
and in ideal cases $\bar{n}_\mathrm{ex}$ can be zero.
But there are often extra photons to be added to Eq.\,(\ref{eq:nav})
in actual experiments, for example
$\bar{n}_\mathrm{ex}\approx 0.062$ in our present experiment.
Such extra photons increases the squeezing level required to surpass
the classical limit from the minimum mentioned above.
% (unless this excess noise is reduced).
Furthermore, the probe beam
to control relative phases among the beams
is present in the signal beam.
This must be removed before the transmission channel
(by using a different wavelength or a different polarization),
and the local ocillators in homodyne detection must be
controlled referring to the signal itself.
These are, however, technical rather than principal issues.
% When all these conditions are satisfied, the amount of information
% carried by dense coding can be that shown in Fig.\,\ref{fg:ixy}.
%%1019

%%%%%%%%%%%%%%%%%%%%%%%%%%%%%%%%%%%%%%%%%%%%%%%%%%%%%%%%%%%%
%
\section{Conclusion}
%
% In conclusion,
We demonstrate the entanglement assisted coding
using a two-mode squeezed vacuum state,
whose capacity exceeds those of conventional coherent state channels
and of squeezed state homodyne channel.
The scheme is useful, even in the present form,
for sensing with weak optical power.
When sufficiently high level of squeezing become available,
quantum dense coding using CV can be realized
as the natural extension of the present scheme.
From the technical aspect, basic operations on CV such as
the displacement of an EPR beam and
the recovery of separable squeezed states from the EPR beams
are demonstrated.

%%%%%%%%%%%%%%%%%%%%%%%%%%%%%%%%%%%%%%%%%%%%%%%%%%%%%%%%%%%%
%
\begin{acknowledgments}
The authors are grateful to Dr.~Ban, Dr.~Aoki, 
Dr.~Fujiwara, Dr.~Takeoka,  % Dr.~Hasegawa, Dr.~Mitsumori, 
Mr.~Takei, Mr.~Hiraoka, and Mr.~Yonezawa 
for their help and discussions.
\end{acknowledgments}
%
%%%%%%%%%%%%%%%%%%%%%%%%%%%%%%%%%%%%%%%%%%%%%%%%%%%%%%%%%%%%
%
%	references
%

%
%%%%%%%%%%%%%%%%%%%%%%%%%%%%%%%%%%%%%%%%%%%%%%%%%%%%%%%%%%%%
%

\begin{thebibliography}
\raggedright
%
\bibitem{FK98}
  A.~Furusawa, J.L.~S\o rensen, S.L.~Braunstein, C.A.~Fuchs,
  H.J.~Kimble, and E.S.~Polzik,
  ``Unconditional Quantum Teleportation,''
  Science \textbf{282}, 706 %706--709
  (1998).
%
\bibitem{ZGCLK03}
  T.C.~Zhang, K.W.~Goh, C.W.~Chou, P.~Lodahl, H.J.~Kimble,
  ``Quantum teleportation of light beams,''
  \pra\textbf{67}, 033802 (2003).
%
\bibitem{BTBSRBSL03}
  W.P.~Bowen, N.~Treps, B.C.~Buchler, R.~Schnabel, T.C.~Ralph,
  H.-A.~Bachor, T.~Symul, P.K.~Lam,
  ``Experimental investigation of continuous variable quantum teleportation,''
  \pra\textbf{67}, 032302 (2003).
%
\bibitem{Ban99}
  M.~Ban,
  ``Quantum Dense Coding via a Two-mode Squeezed-vacuum State,''
  J.~Opt.~B \textbf{1}, L9
  (1999).
%
\bibitem{Ban00}
  M.~Ban,
  ``Information transmission via dense coding in a noisy quantum channel,''
  Phys.~Lett.\ A \textbf{276}, 213 %213--220
  (2000).
%
\bibitem{BK00}
  S.L.~Braunstein and H.J.~Kimble,
  ``Dense coding for continuous variables,''
  \pra\textbf{61}, 042302 (2000).
%
\bibitem{LJZXP02}
  X.~Li, Q.~Pan, J.~Jing, J.~Zhang, C.~Xie, and K.~Peng,
  ``Quantum Dense Coding Exploiting a Bright Einstein-Podolsky-Rosen Beam,''
  \prl\textbf{88}, 047904 (2002).
%
\bibitem{RH02}
  T.C.~Ralph and E.H.~Huntington,
  ``Unconditional continuous-variable dense coding,''
  \pra\textbf{66}, 042321 (2002).
%
\bibitem{SB98a}
  S.L.~Braunstein,
  ``Error Correction for Continuous Quantum Variables,''
  \prl\textbf{80}, 4084 %4084--4087
  (1998).
%
\bibitem{SB98b}
  S.L.~Braunstein,
  ``Quantum error correction for communication with linear optics,''
  \nat\textbf{394}, 47 %47--49
  (1998).
%
\bibitem{LloydBraunstein99}
  S.~Lloyd and S.L.~Braunstein,
  ``Quantum Computation over Continuous Variables,''
  \prl\textbf{82}, 1784 (1999).
%
\bibitem{GKP01}
  D.~Gottesman, A.~Kitaev, and J.~Preskill,
  ``Encoding a qubit in an oscillator,''
  \pra\textbf{64}, 012310 (2001).
%
\bibitem{BS02}
   S.D.~Bartlett and B.C.~Sanders,
  ``Universal continuous-variable quantum computation:
    Requirement of optical nonlinearity for photon counting,''
  \pra\textbf{65}, 042304 (2002).
%
\bibitem{PCK92}
  E.S.~Polzik, J.~Carry, and H.J.~Kimble,
  ``Atomic Spectroscopy with Squeezed Light
    for Sensitivity Beyond the Vacuum-State Limit,''
  \ap\textbf{B55}, 279 %279--290
  (1992).
%
\end{thebibliography}
\end{document}